\documentclass[11pt]{article}
\usepackage{epsfig}

\newcommand{\BABARPubYear}    {08}

\newcommand{\BABARConfNumber} {019}
\newcommand{\SLACPubNumber} {13328}

\input babarsym
\def\cpoddhiggs      {\ensuremath{{A^{0}}}\xspace}

\def\n1Spipi     {\ensuremath{}\xspace}

\def\gamgam      {\ensuremath{e^+e^-\to\gamma\gamma}\xspace}
\def\bhabha      {\ensuremath{e^+e^-\to e^+e^-\gamma}\xspace}

\def\beq{\begin{equation}}
\def\eeq{\end{equation}}
\def\bea{\begin{eqnarray}}
\def\eea{\end{eqnarray}}
\def\bq{\begin{quote}}
\def\eq{\end{quote}}
\def\bi{\begin{itemize}}
\def\ei{\end{itemize}}
\def\bc{\begin{center}}
\def\ec{\end{center}}

\def\etal{{\em et al.}}

\long\def\inst#1{\par\nobreak\kern 4pt\nobreak
    {\it #1}\par\vskip 10pt plus 3pt minus 3pt}

\begin{document}
{\pagestyle{empty}

\begin{flushright}
\babar-CONF-\BABARPubYear/\BABARConfNumber \\
%\babar-PUB-\BABARPubYear/\BABARPubNumber \\
SLAC-PUB-\SLACPubNumber \\
%hep-ex/\LANLNumber \\
July 2008 \\
\end{flushright}

\par\vskip 5cm

% Title of the paper
\begin{center}
\Large \bf 
Search for Invisible Decays of a Light Scalar in Radiative
Transitions $\Upsilon(3S)\to\gamma\cpoddhiggs$
\end{center}
\bigskip

\begin{center}
\large The \babar\ Collaboration\\
\mbox{ }\\
July 31, 2008
\end{center}
\bigskip \bigskip

% Abstract
\begin{center}
\large \bf Abstract
\end{center}
We search for a light scalar particle produced in single-photon decays
of the $\Upsilon(3S)$ resonance through the process
$\Upsilon(3S)\to\gamma+A^0$,
$A^0\to\mathrm{invisible}$. Such an object 
appears in Next-to-Minimal Supersymmetric extensions of the Standard
Model, where a light \CP-odd Higgs boson naturally couples strongly to
$b$-quarks. If, in addition, there exists a light, stable neutralino,
decays of $A^0$ could be preferentially to an invisible final
state. We search for events with a single high-energy photon and a
large missing mass, consistent with a 2-body decay of $\Upsilon(3S)$. 
We find no evidence for such processes in 
a sample of $122\times10^6$ $\Upsilon(3S)$  decays collected
by the \babar\ collaboration at the \pep2\ B-factory, and set 90\%
C.L. upper limits 
on the branching fraction
$\mathcal{B}(\Upsilon(3S)\to\gamma A^0)\times\mathcal{B}(A^0\to\mathrm{invisible})$ at
$(0.7-31)\times10^{-6}$ in the mass range $m_{A^0}\le7.8$~GeV. The
results are preliminary. 

\vfill
\begin{center}

Submitted to the 34$^{\rm th}$ International Conference on High-Energy Physics, ICHEP 08,\\
29 July---5 August 2008, Philadelphia, Pennsylvania.

\end{center}

\vspace{1.0cm}
\begin{center}
{\em Stanford Linear Accelerator Center, Stanford University, 
Stanford, CA 94309} \\ \vspace{0.1cm}\hrule\vspace{0.1cm}
Work supported in part by Department of Energy contract DE-AC02-76SF00515.
\end{center}

\newpage
} % end of pagestyle{empty}

% Input author list file
%
%author list removed temporarily to save trees 7/9/04 RNC
%
\begin{center}
\small

The \babar\ Collaboration,
\bigskip

%% author list as of 02-Jul-2008 (523 authors)
%
B.~Aubert,
M.~Bona,
Y.~Karyotakis,
J.~P.~Lees,
V.~Poireau,
E.~Prencipe,
X.~Prudent,
V.~Tisserand
\inst{Laboratoire de Physique des Particules, IN2P3/CNRS et Universit\'e de Savoie, F-74941 Annecy-Le-Vieux, France }
J.~Garra~Tico,
E.~Grauges
\inst{Universitat de Barcelona, Facultat de Fisica, Departament ECM, E-08028 Barcelona, Spain }
% ...

\end{center}\newpage

% The body of the paper starts here
\section{INTRODUCTION}
\label{sec:Introduction}

The search for the origin of mass is one of the great quests in
particle physics. Within the Standard Model~\cite{ref:SM}, 
fermion and gauge boson 
masses are generated by the Higgs mechanism through the spontaneous
breaking of the electroweak symmetry. A single Standard Model Higgs
boson is required to be heavy, with the mass constrained by 
direct searches to $m_{H} >114.4$~GeV \cite{Barate:2003sz}, and by
precision electroweak measurements to 
$m_{H} = 129^{+74}_{-49}$~GeV \cite{LEPSLC:2005ema}.

The Standard Model and the simplest electroweak symmetry breaking
scenario suffer from quadratic divergences in the radiative
corrections to the mass parameter of the Higgs potential. 
Several theories beyond the Standard Model that regulate these
divergences have been proposed. 
Supersymmetry~\cite{ref:SUSY} is one such model; however, in its simplest form 
(the Minimal Supersymmetric Standard Model, MSSM) questions of
parameter fine-tuning and ``naturalness'' of the Higgs mass scale
remain. 

Theoretical efforts to solve unattractive features of MSSM often
result in models that introduce additional Higgs fields, with one of
them naturally light. For instance, the Next-to-Minimal Supersymmetric
Standard Model (NMSSM)~\cite{Dermisek:2005ar} introduces a singlet
Higgs field. A linear combination of this singlet state with a member
of the electroweak doublet produces a \CP-odd Higgs state $A^0$ 
whose mass is not required to be large. 
Direct searches typically constrain $m(A^0)$ to be
below $2m_b$~\cite{Dermisek:2006} making it accessible to decays of
$\Upsilon$ resonances. 
An ideal place to search for such \CP-odd Higgs would be
$\Upsilon \to \gamma \cpoddhiggs$, as originally proposed by Wilczek
\cite{Wilczek:1977zn}.  A study of the NMSSM parameter
space \cite{Dermisek:2006py} predicts the
branching fraction to this final state to be as high as $10^{-4}$. 

The decays of the light Higgs boson depend on its
mass and couplings, as well as on the low-energy particle spectrum of
the underlying theory. In certain NMSSM scenarios, particularly those
in which the mass of the lightest  
supersymmetric particle (LSP) is above $m_\tau$ or if 
$m_{\cpoddhiggs} < 2m_{\tau}$, the dominant decay mode of
$\cpoddhiggs$ may be invisible: $\cpoddhiggs\to\chi^0\bar{\chi}^0$, where
the neutralino $\chi^0$ is the LSP. The
cleanest experimental signature of such decays is production of 
monochromatic single photons in decays $\Upsilon\to\gamma\cpoddhiggs$,
accompanied by a significant missing energy and momentum. 
The photon energy in the $\Upsilon$ center-of-mass
(CM)~\footnote{Hereafter $*$ denotes a CM quantity} is given by 
\begin{equation}
E_\gamma^{*} = \frac{m_{\Upsilon}^2-m_{\cpoddhiggs}^2}{2m_{\Upsilon}}\ .
\label{eq:Egamma}
\end{equation}

The current best limit on the branching fraction 
$\mathcal{B}(\Upsilon\to\gamma X)$ with $X\to\mathrm{invisible}$ comes
from a measurement by the CLEO collaboration on
$\Upsilon(1S)$~\cite{CLEO:1994ch}. The quoted limits range from
$1.3\times10^{-5}$ for the lightest $m_X$ (highest-energy photons) to
$(4$--$6)\times10^{-4}$ 
for $m_X\approx 8$~GeV (PDG quotes this result as 
$\mathcal{B}(\Upsilon(1S)\to\gamma X)<3\times10^{-5}$ for
$m_X<7.2$~GeV \cite{PDBook}). There are currently no competitive
measurements at the higher-mass $\Upsilon$ resonances. 

In the following, we describe a search for a monochromatic peak in the
missing mass distribution of events with a single high-energy
photon. We assume that the decay width of \cpoddhiggs is negligibly
small compared to experimental resolution, as
expected~\cite{ref:Lozano} for $m_{\cpoddhiggs}$ sufficiently
far from the mass of $\eta_b$~\cite{ref:etab}. Furthermore, we assume
that a single \cpoddhiggs state exists in the range
$0<m_\cpoddhiggs\le7.8$~GeV; or if two or more states are present,
they do not interfere.  

\section{THE \babar\ DETECTOR AND DATASET}
\label{sec:babar}

We search for
two-body transitions $\Upsilon(3S)\to\gamma\cpoddhiggs$, followed by invisible
decays of $\cpoddhiggs$ in a sample of $(121.8\pm 1.2)\times10^6$
$\Upsilon(3S)$ 
decays collected with the \babar\ detector
at the \pep2\ asymmetric-energy \epem\ collider at the Stanford Linear
Accelerator Center. The data were
collected at the nominal CM energy $E_{cm}=10.355$~GeV.
The CM frame was boosted relative to the
detector approximately along the detector's magnetic field axis by
$\beta_z=0.469$.  

For characterization of the background events
we also use a sample of
$0.97~\mathrm{fb}^{-1}$ collected $30$~MeV below the \Y2S
resonance, a sample of $2.6~\mathrm{fb}^{-1}$ collected $30$~MeV below
the \Y3S resonance, 
$\Y4S$ decays corresponding to the integrated luminosity of
$4.7~\mathrm{fb}^{-1}$, and $4.5~\mathrm{fb}^{-1}$ integrated above
the \Y4S resonance. We henceforth refer to these datasets as the 
{\em off-resonance\/} sample. 

Since the \babar\ detector is described in detail
elsewhere~\cite{detector}, 
only the components of the detector crucial to this analysis are
 summarized below. 
Charged particle tracking is provided by a five-layer double-sided silicon
vertex tracker (SVT) and a 40-layer drift chamber (DCH). 
Photons and neutral pions are identified and measured using
the electromagnetic calorimeter (EMC), which comprises 6580 thallium-doped CsI
crystals. These systems are mounted inside a 1.5-T solenoidal
superconducting magnet. 
The Instrumented Flux Return (IFR) forms the return yoke of
the superconducting coil, instrumented in the central barrel region with
limited streamer tubes for the identification 
of muons and the detection of clusters produced
by neutral hadrons. 
We use the GEANT~\cite{geant} software to simulate interactions of particles
traversing the \babar\ detector, taking into account the varying
detector conditions and beam backgrounds. 

\section{SINGLE PHOTON TRIGGER}
\label{sec:trigger}

Detection of the low-multiplicity single photon events requires 
dedicated trigger and filter lines. Event processing and selection
proceeds in three steps. First, the hardware-based Level-1 (L1) trigger accepts
single-photon events
if they contain at least one EMC cluster with energy above $800$~MeV
(in the laboratory frame). The total L1 trigger rate was typically 4--5 kHz
for a combination of 24 trigger topologies,
including the single-photon line which contributed the rate of
300--400 Hz. 
Second, L1-accepted events are forwarded to a
software-based Level-3 (L3) trigger, which forms DCH tracks and EMC
clusters and makes decisions for a variety of physics signatures. Two
single-photon L3 trigger lines were active during the data taking 
period. The high-energy (``HighE'') line requires an isolated EMC
cluster with CM energy $E^{*}_\gamma>2$~GeV, and no tracks originating 
from the $e^+e^-$ interaction region. A subset of the data, amounting
to $(82.8\pm0.8)\times10^6$ \Y3S decays and $2.6~\mathrm{fb}^{-1}$
collected 30~MeV below the \Y3S, were also processed with a low-energy
(``LowE'') single-photon trigger, which requires an EMC cluster with 
CM energy $E^{*}_\gamma>1$~GeV, and no tracks originating 
from the $e^+e^-$ interaction region. The acceptance rate of the two
single-photon L3 lines was up to 100 Hz. Events accepted by L3 are
written to mass storage, at the rate of up to 900~Hz. 

Additional requirements are applied to the events at the
reconstruction stage. We process single-photon events if they satisfy
one of the two criteria. The ``HighE'' selection requires one EMC cluster
in the event with a CM energy $E^{*}_\gamma>3$~GeV and no DCH tracks
with momentum $p^{*}>1$~GeV. The ``LowE'' selection requires 
one EMC cluster with the transverse profile consistent with an
electromagnetic shower and a CM energy $E^{*}_\gamma>1.5$~GeV, and no
DCH tracks with momentum $p^{*}>0.1$~GeV. The two selection criteria
are not mutually exclusive. 

\section{EVENT SELECTION AND YIELDS}
\label{sec:Analysis}

Due to the specifics of the online and reconstruction selections, we 
split the dataset into two broad energy ranges based on the energy of
the highest-energy (in the CM frame) EMC cluster. 
The high-energy region, accepted by ``HighE'' L3 and reconstruction
selections, corresponds to $3.2<E^{*}_\gamma<5.5$~GeV. The
backgrounds in this region are 
dominated by the QED process $e^+e^-\to\gamma\gamma$, especially near 
$E^{*}_\gamma=E_{cm}/2$, where the photon energy distribution for
\gamgam events  
peaks. The offline event selection is optimized to reduce this peaking
background as much as possible. 

The second energy range is $2.2<E^{*}_\gamma<3.7$ GeV, which
corresponds to the ``LowE'' online selection. This region is dominated
by the low-angle radiative Bhabha events $e^+e^-\to e^+e^-\gamma$, in which
both electron and positron miss the sensitive detector volumes. In the 
region $3.0<E^{*}_\gamma<3.7$ GeV, the tail from the
$e^+e^-\to\gamma\gamma$ background is significant. 

A limited number of variables are available for these very
low-multiplicity event samples. We use the following 
variables to select the events of interest: 
\bi
\item Photon quality: number of crystals in the EMC cluster
  $N_\mathrm{crys}$, and transverse shower moments $LAT$ and
  $a_{42}$~\cite{ref:LAT}. 
\item Fiducial selection of the primary photons: cosine of the CM polar
  angle $\cos\theta^{*}_\gamma$ and the azimuthal angle 
  $\phi^{*}_\gamma$. The signal photons are expected to be distributed
  as $1+\cos^2\theta^{*}$, while the backgrounds are more strongly
  peaked in the forward and backward directions. 
\item Extra particles in the event: we require that no charged-particle
  tracks are found
  in the SVT and the DCH. We also apply cuts on the energy of the
  second-highest photon in the event $E^{*}_2$ (computed in CM frame),
  extra energy in the calorimeter
  $E_\mathrm{extra}=E_\mathrm{total}-E_\gamma$, computed in the lab frame,
  and the azimuthal angle difference between the primary and the second
  photon in the   event $\phi^{*}_2-\phi^{*}_1$. 
  Non-zero $E_\mathrm{extra}$ may be present in the signal events due
  to machine backgrounds.
  The cut on $\phi^{*}_2-\phi^{*}_1$ suppresses \gamgam
  and other QED backgrounds.
\item IFR veto: we cut on the azimuthal angle difference between the
  primary photon and any IFR cluster. This variable,
  $\Delta\phi^{*}_\mathrm{NH}$, rejects the $e^+e^-\to\gamma\gamma$ events
  in which one of the photons is lost in the dead regions between the
  EMC crystals, but is reconstructed as an IFR cluster. 
\ei

We optimize the event selection to maximize
$\varepsilon_S/\sqrt{\varepsilon_B}$, where $\varepsilon_S$ is the
selection efficiency for the signal, and $\varepsilon_B$ is the
background efficiency. We use Monte Carlo 
samples generated over a broad range $0<m_\cpoddhiggs\le8$~GeV of
possible $A^0$ masses for the 
signal events. We also use approximately 10\% of the available dataset
as a background sample for the selection optimization. This sample is
included in the final fit.

In the following, we present the analysis of the data in each energy
range separately. We use the high-energy region to measure the signal
yields in the mass range $0<m_\cpoddhiggs\le6$~GeV. We measure the
yields in the region $6<m_\cpoddhiggs\le 7.8$~GeV using the
low-energy region. The overlap between the two regions is minimal, and
the events yields are consistent in the range of $m_\cpoddhiggs$ where
the regions overlap. 

\subsection{HIGH-ENERGY REGION}
\label{sec:highE}

The final selection for the energy range $3.2<E^{*}_\gamma<5.5$~GeV is
summarized in Table~\ref{tab:cuts}. The selection efficiency
for signal is 10-11\%, depending on $m_\cpoddhiggs$, and is below
$10^{-5}$ for \gamgam events. Most of the signal efficiency loss
occurs due to the fiducial requirements: the CM polar angle selection 
$-0.31< \cos\theta^{*}_\gamma < 0.6$ ensures that the second photon from
\gamgam background would hit the 
barrel regions of the EMC and the IFR, and the azimuthal requirement
$\cos(6\phi^{*}_\gamma)<0.96$ vetoes the dead regions between the six
equal-size IFR sectors. 
\begin{table}
\caption{Selection criteria for the two regions, low and high energies.
}
\label{tab:cuts}
\bc
\begin{tabular}{lcc}
\hline\hline
Variable & $3.2<E^{*}_{\gamma}<5.5$~GeV & $2.2<E^{*}_{\gamma}<3.7$~GeV \\
\hline
Number of crystals in EMC cluster & $20<N_\mathrm{crys}<48$ & $12<N_\mathrm{crys}<36$\\
LAT shower shape & $0.24 < LAT < 0.51$ & $0.15 < LAT < 0.49$ \\
$a_{42}$ shower shape & $a_{42}<0.07$ & $a_{42}<0.07$ \\
Polar angle acceptance & $-0.31< \cos\theta^{*}_\gamma < 0.6$ & $-0.46< \cos\theta^{*}_\gamma < 0.46$ \\
2nd highest cluster energy (CMS) & $E^{*}_2<0.2$~GeV & $E^{*}_2<0.14$~GeV \\
Extra photon correlation & $\cos(\phi^{*}_2-\phi^{*}_1)>-0.95$ & $\cos(\phi^{*}_2-\phi^{*}_1)>-0.95$ \\
Extra EMC energy (Lab) & $E_\mathrm{extra} < 0.1$~GeV & $E_\mathrm{extra} < 0.22$~GeV \\
IFR veto & $\cos(\Delta\phi^{*}_\mathrm{NH}) > -0.9$ & $\cos(\Delta\phi^{*}_\mathrm{NH}) > -0.95$ \\\
IFR fiducial & $\cos(6\phi^{*}_\gamma)<0.96$ & ...\\
\hline\hline
\end{tabular}
\ec
\end{table}

We extract the yield of signal events as a function of the assumed
mass $m_\cpoddhiggs$ in the interval $0<m_\cpoddhiggs\le 6$~GeV by
performing a series of unbinned extended maximum likelihood fits
to the distribution of the missing mass squared
\beq
m_X^2 \equiv m_{\Y3S}^2 - 2E^{*}_\gamma m_{\Y3S} 
\label{eq:mX2}
\eeq
in fine steps of $\Delta m_\cpoddhiggs=0.1$~GeV. 
After the final selection, 955 events remain in the data sample in
the interval $-5\le m_X^2\le 40~\mathrm{GeV}^2$. 
The dominant
background in this region is from \gamgam, radiative Bhabha, and
two-photon fusion events. The background from \gamgam is particularly
problematic, since its distribution peaks near $m_X^2=0$. We determine
the probability density function (PDF) of this background by selecting
a high-statistics sample of on-resonance \gamgam events with the IFR
veto removed (a total of 244,462 events).  We determine the
efficiency of 
the IFR veto by selecting \gamgam events with one photon in the
off-resonance sample. We 
find 
$\varepsilon_\mathrm{IFR\,veto}=
N_{\gamma\gamma\,\mathrm{veto}}/N_{\gamma\gamma\,\mathrm{no\,veto}} = 
(4.5\pm1.9)\times 10^{-4}$, where
$N_{\gamma\gamma\,\mathrm{veto}}$ is the number of events accepted
with the full selection (Table~\ref{tab:cuts}), and 
$N_{\gamma\gamma\,\mathrm{no\,veto}}$ is the number of events accepted
with the same selection but IFR veto removed. 
The uncertainty accounts for the time-dependent variation in
$\varepsilon_\mathrm{IFR\,veto}$ between the different off-resonance
samples. We then fix the number of expected \gamgam events to 
$N_{\gamma\gamma} = 110\pm46$. 

The background from the radiative Bhabha and two-photon fusion events is
described by a smooth exponential function 
$f_\mathrm{bkg}(m_X^2)\propto \exp(c m_X^2)$. The parameter $c$ and
the yield of this continuum background are left free in the fit. 
\begin{figure}[t!]
\bc
\epsfig{file=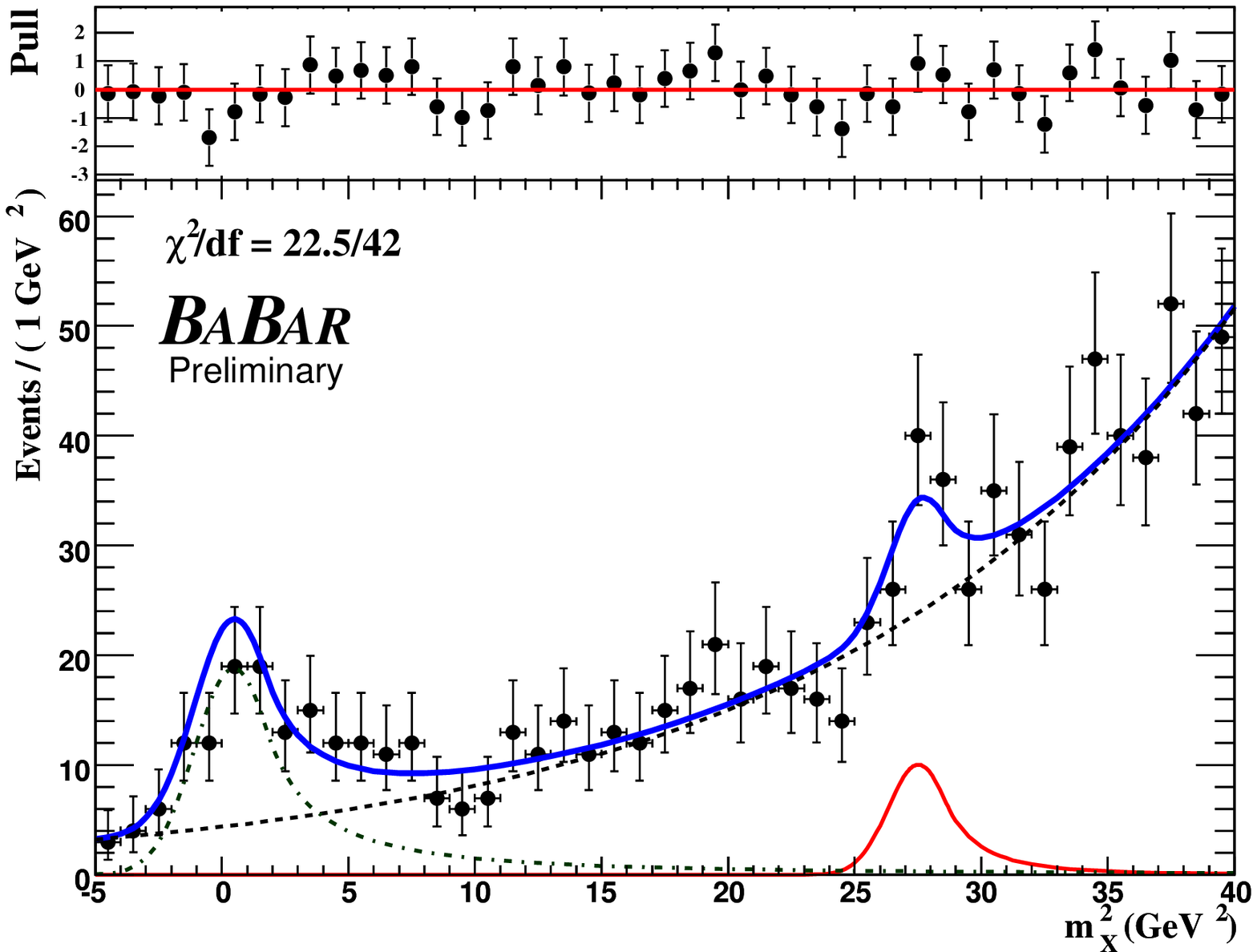,width=4.4in} 
\vspace{-\baselineskip}
\ec
\caption{Sample fit to the high-energy dataset ($122\times10^6$ \Y3S
  decays). The bottom plot shows the data (solid 
  points)   overlaid by 
  the full PDF curve (solid blue line), signal contribution with
  $m_\cpoddhiggs=5.2$~GeV (solid red line),
  \gamgam contribution   (dot-dashed green line), and continuum
  background PDF (black dashed 
  line). The top plot shows the pulls
  $p=(\mathrm{data}-\mathrm{fit})/\sigma(\mathrm{data})$  with unit
  error bars.}
\label{fig:fit52_float}
\end{figure}
\begin{figure}[h!]
\bc
\epsfig{file=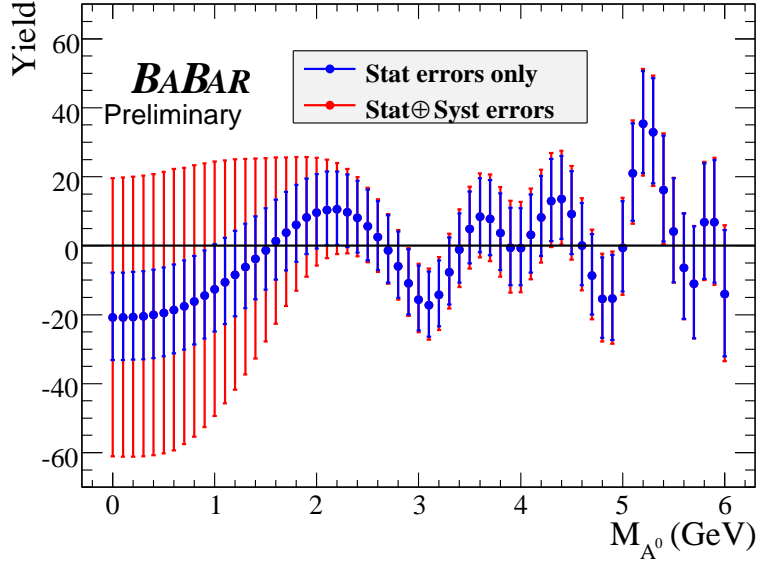,width=4.4in} 
\vspace{-\baselineskip}
\ec
\caption{Signal yields $N_\mathrm{sig}$ as a function of assumed mass
  $\m_{\cpoddhiggs}$ in the high-energy dataset.
  Blue error bars are statistical only, and the red error bars include
  the systematic contributions.  Since the spacing between the points
  is smaller than the experimental resolution, the neighboring points
  are highly correlated. 
}
\label{fig:scanHighE}
\end{figure}

Our Monte Carlo simulations estimate that the backgrounds from the
generic \Y3S decays or misreconstructed vector mesons produced through
initial-state radiation (ISR) processes are negligible. The ISR
processes can potentially contribute peaking backgrounds at low
$m_X^2$. We see no evidence for these extra contributions in the
off-resonance sample, but also vary the peaking \gamgam PDF to
estimate potential systematic effects. 

The signal PDF is described by a Crystal Ball~\cite{ref:CBshape}
function centered around the expected value of
$m_X^2=m_{\cpoddhiggs}^2$. We determine the PDF as a function of
$m_{\cpoddhiggs}$ using high-statistics simulated samples of signal
events, and we determine the uncertainty in the PDF parameters by
comparing the distributions of the simulated and reconstructed \gamgam
events. The resolution for signal events varies between
$\sigma(m_X^2)=1.5~\mathrm{GeV}^2$ for $m_\cpoddhiggs\approx 0$ to 
$\sigma(m_X^2)=0.7~\mathrm{GeV}^2$ for $m_\cpoddhiggs=8$~GeV. 

A sample fit, for $m_\cpoddhiggs=5.2$~GeV, is shown in
Fig.~\ref{fig:fit52_float}. This fit corresponds to the signal yield
of $N_\mathrm{sig}=37\pm15$, with the statistical significance of 
$\mathcal{S}\equiv \sqrt{2\ln(L_{\max}/L_0)} = 2.6\sigma$, where
$L_{\max}$ is the maximum value of the likelihood, and $L_0$ is the
value of the likelihood with the signal yield fixed to zero. No other
values of $m_\cpoddhiggs<6$~GeV return higher significance or
likelihood. The results of the fits in fine steps of $m_\cpoddhiggs$
are shown in Fig.~\ref{fig:scanHighE}. 

%%%%%%%%%%%%%%%%%%%%%%%%%%%%%%%%%%%%%%%%
\subsection{LOW-ENERGY REGION}
\label{sec:lowE}

The final selection for the energy range $2.2<E^{*}_\gamma<3.7$~GeV is
summarized in Table~\ref{tab:cuts}. The selection efficiency
for signal is 20\%. Most of the signal efficiency loss
occurs due to the fiducial requirement on the CM polar angle
$|\cos\theta^{*}_\gamma|<0.46$, applied to suppress the background from
\bhabha, which rises steeply in the forward and backward
directions. We restrict the photon energy range to avoid the region
$E^{*}_\gamma<2.2$~GeV where the backgrounds are excessively
high and the single-photon trigger selection requires further investigation. 

We extract the yield of the signal events as a function of the assumed
mass $m_\cpoddhiggs$ in the range $6<m_\cpoddhiggs\le7.8$~GeV by
performing a set of unbinned extended maximum likelihood fits 
to the distribution of the missing mass squared $m_X^2$ in steps of
$\Delta m_\cpoddhiggs=0.025$~GeV.  
After the final selection, 14,947 events remain in the data sample in
the interval $30\le m_X^2\le 62~\mathrm{GeV}^2$. 
The dominant
background in this region is from the radiative Bhabha \bhabha events,
with contributions from \gamgam becoming
relevant at low values of $m_X^2$ (high photon energy). 

We parameterize the background from the radiative Bhabha events 
by a smooth exponential function 
$f_\mathrm{Bhabha}(m_X^2)\propto \exp(c_1 m_X^2+c_2 m_X^4)$. The
parameters $c_1$ and $c_2$, as well as the yield of \bhabha events are
left free in the fit. This PDF also accounts for other radiative
processes, such as $e^+e^-\to\tau^+\tau^-\gamma$ and
$e^+e^-\to\mu^+\mu^-\gamma$. 

The \gamgam distribution is modeled by a sample of simulated
events with an equivalent integrated luminosity of
$22~\mathrm{fb}^{-1}$. The PDF has two components: 
a smooth continuum 
with a turn-on of $e^+e^-\to3\gamma$ events starting around
$m_X^2=53~\mathrm{GeV}^2$, and a broad peak with
the width of about $2.5~\mathrm{GeV}^2$ which accounts for 
a forward and backward corner of $3\gamma$ phase space. 
The fraction of this peak, the fraction of the continuum
$e^+e^-\to3\gamma$ events, and the normalization of the $\gamgam(\gamma)$
events are left free in the fit. 
The signal PDF is described by the same Crystal Ball~\cite{ref:CBshape}
function as in the high-energy region. 

\begin{figure}[ht!]
\bc
\epsfig{file=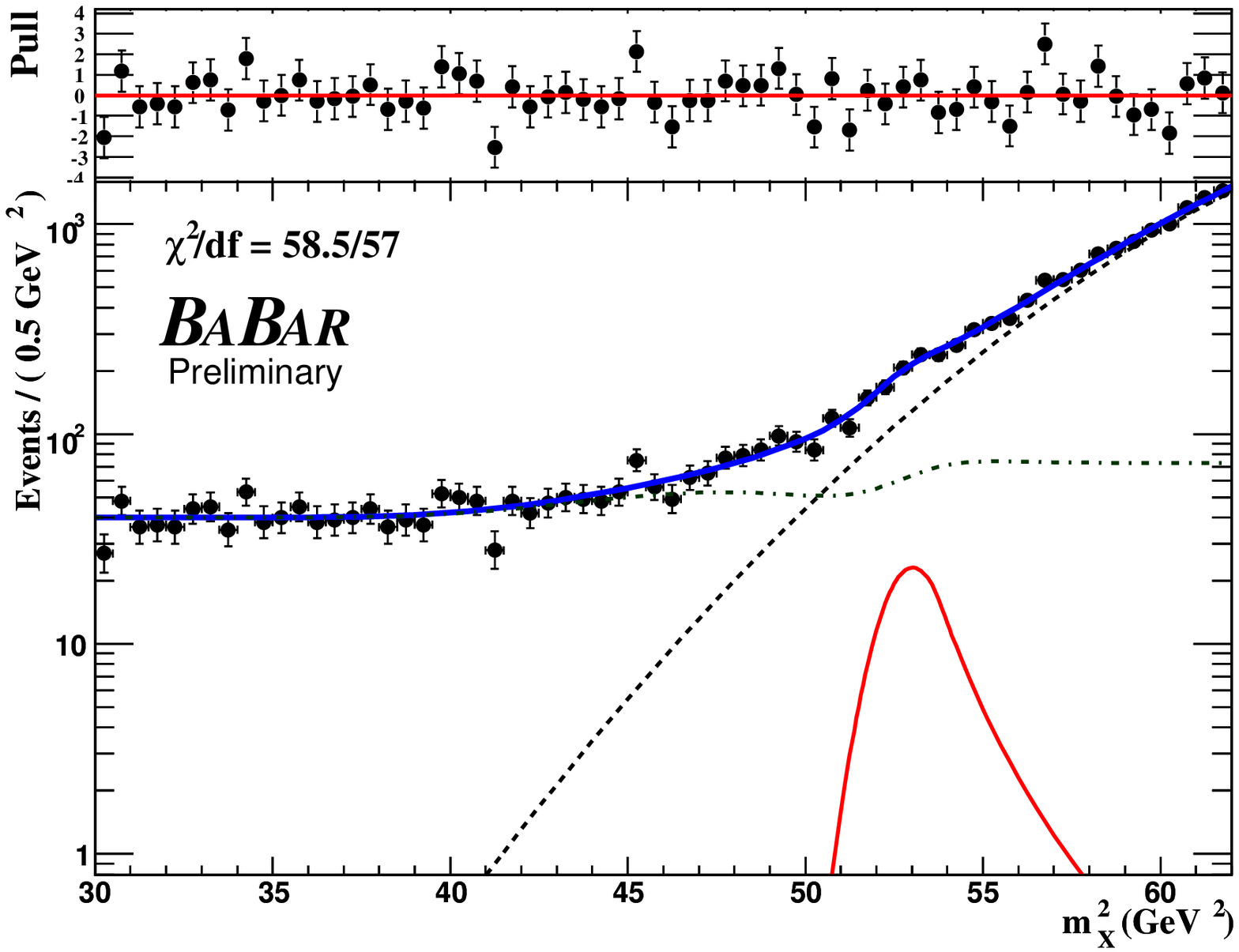,width=4.4in} 
\vspace{-\baselineskip}
\ec
\caption{Sample fit to the low-energy dataset ($83\times10^6$ \Y3S
  decays). The bottom plot shows the data (solid 
  points)   overlaid by 
  the full PDF curve (solid blue line), signal contribution with
  $m_\cpoddhiggs=7.275$~GeV (solid red line),
  \gamgam contribution   (dot-dashed green line), and continuum
  background PDF (black dashed 
  line). The top plot shows the pulls
  $p=(\mathrm{data}-\mathrm{fit})/\sigma(\mathrm{data})$ with unit
  error bars.} 
\label{fig:fit1dGGfull_73}
\end{figure}

\begin{figure}[ht!]
\bc
\epsfig{file=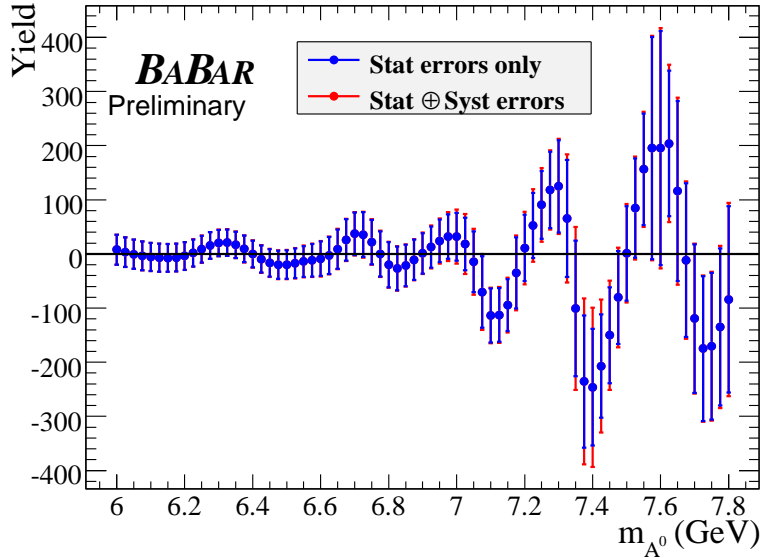,width=4.4in} 
\vspace{-\baselineskip}
\ec
\caption{Signal yields $N_\mathrm{sig}$ as a function of assumed mass
  $\m_{\cpoddhiggs}$ in the ``LowE'' region.
  Blue error bars are statistical only, and the red error bars include
  the systematic contributions. Since the spacing between the points
  is smaller than the experimental resolution, the neighboring points
  are highly correlated. 
}
\label{fig:fullResultsLowE}
\end{figure}

A fit for $m_\cpoddhiggs=7.275$~GeV, is shown in
Fig.~\ref{fig:fit1dGGfull_73}. This fit corresponds to the signal yield
of $N_\mathrm{sig}=119\pm71$, with statistical significance of 
$\mathcal{S}\equiv \sqrt{2\ln(L_{\max}/L_0)} = 1.7\sigma$. 
No other
values in the range $6<m_\cpoddhiggs\le 7.8$~GeV return higher
significance. This fit also returns 
$N_\mathrm{Bhabha}=11419\pm441$ and $N_{\gamma\gamma}=3410\pm451$.
The results of the fits in fine steps of $m_\cpoddhiggs$ 
are shown in Fig.~\ref{fig:fullResultsLowE}. 
For each fit where the signal yield is
allowed to vary, we also allow for the variation of the background
shape parameters. We find that the shapes of the background PDFs are
independent of the assumed $m_\cpoddhiggs$, and is also consistent
with the distribution in the off-resonance sample. 
We observe no significant excess of events
over the range $6<m_\cpoddhiggs\le 7.8$~GeV. 
%

%%%%%%%%%%%%%%%%%%%%%%%%%%%%%%%%%%%%%%%%

\section{SYSTEMATIC UNCERTAINTIES}
\label{sec:Systematics}

The largest systematic uncertainties in the signal yield come from the
estimate of the \gamgam peaking background yield in the high-energy region
and its shape (in both energy regions). Varying the peaking \gamgam
background contribution by its uncertainty changes the signal yield by
$\pm38$ events for $m_\cpoddhiggs=0$, with the effect decreasing with
increased $m_\cpoddhiggs$. The uncertainty due to the \gamgam PDF is
largest in the low-energy region, where it contributes up to $\pm70$
events (for $m_\cpoddhiggs=7.4$~GeV) to the uncertainty in the signal
yield. 

We determine the uncertainty in the signal PDF by comparing the data
and simulated distributions of \gamgam events. We correct for the
differences observed, and use half of the correction as an estimate of
the systematic uncertainty. The effect on the signal yield is
generally small, except for the region near $m_\cpoddhiggs=7.4$~GeV,
where the systematic variation of the signal PDF changes the yield by
$\pm64$ events. Such large variation is caused by high correlation
with the \gamgam yield in this region.
The total additive systematic uncertainty on the yield is ranges
between $1$ and $100$ events, depending on $m_\cpoddhiggs$. 

We measure the trigger and filter selection efficiency using
single-photon \gamgam and \bhabha events selected from a sample of
unbiased randomly accepted triggers. We find excellent agreement with
the Monte Carlo estimates of the trigger efficiency, within the
systematic uncertainty of $0.4\%$. We measure the efficiency of single
photon reconstruction in a large sample of 
$e^+e^-\to\mu^+\mu^-\gamma$, $e^+e^-\to\tau^+\tau^-\gamma$, and 
$e^+e^-\to\gamma\omega$ events, and assign a systematic uncertainty
on the reconstruction efficiency of 2\%. We assign an additional 2\%
systematic uncertainty on the single photon selection. The uncertainty
on the total number of recorded \Y3S decays is estimated to be 1.1\%. 
The total multiplicative error on the branching fraction is $3.1\%$. 

\section{RESULTS AND CONCLUSIONS}
\label{sec:Results}

\begin{figure}[t!]
\bc
\epsfig{file=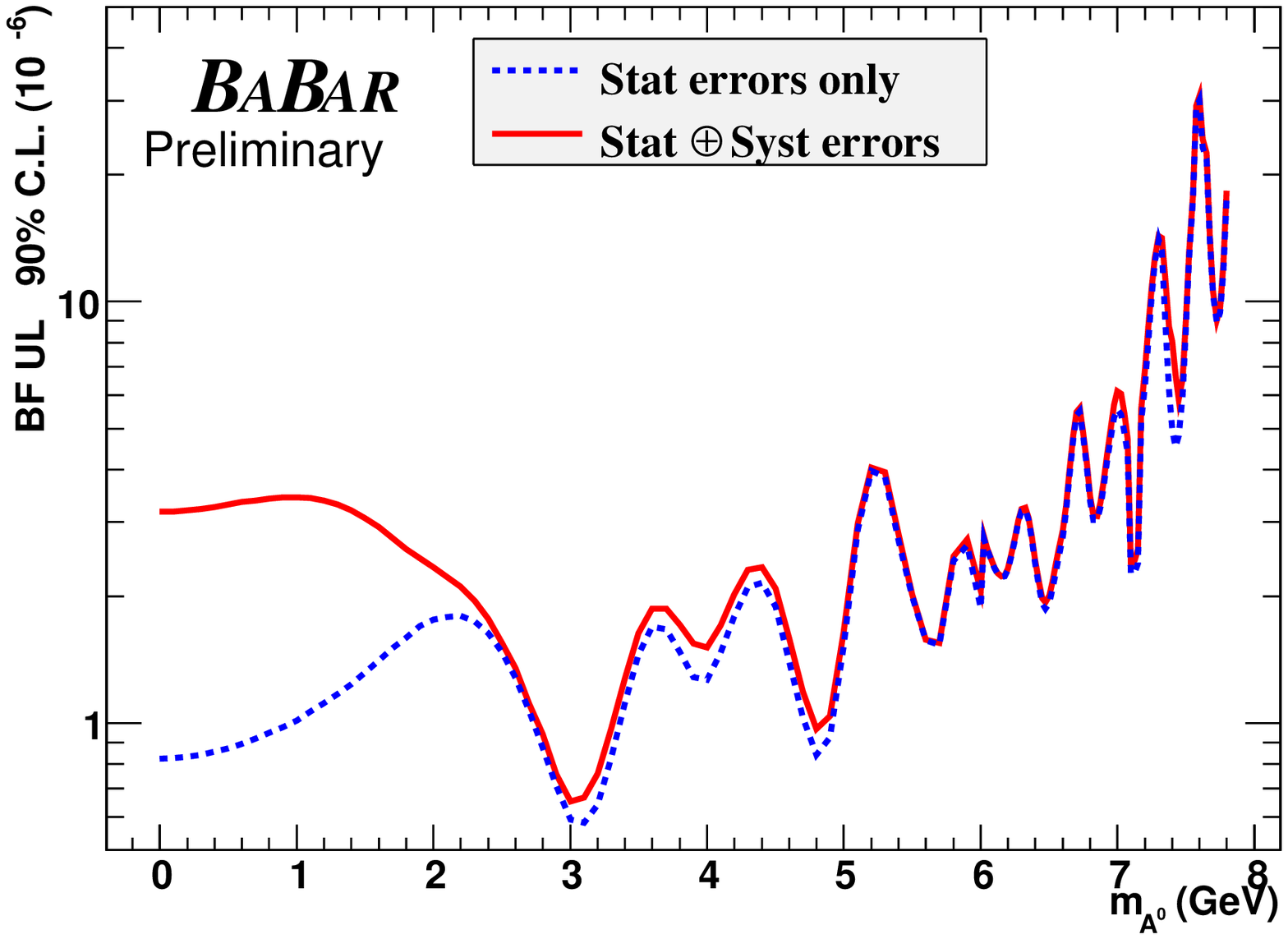,width=6in} 
\ec
\caption{90\% C.L. upper limits on 
the branching fraction
$\mathcal{B}(\Y3S\to\gamma\cpoddhiggs)\times\mathcal{B}(\cpoddhiggs\to\mathrm{invisible})$. 
The dashed blue line shows the statistical uncertainties
  only, the solid red line includes the systematic uncertainties.} 
\label{fig:limitComb}
\end{figure}

We do not observe a significant excess of events above the background
in the range $0<m_\cpoddhiggs\le 7.8$~GeV, and set 
upper limits on the branching fraction
$\mathcal{B}(\Y3S\to\gamma\cpoddhiggs)\times\mathcal{B}(\cpoddhiggs\to\mathrm{invisible})$.
We add statistical and systematic uncertainties (which include the
additive errors on the signal yield and multiplicative uncertainties
on the signal efficiency and the number of recorded \Y3S decays) in
quadrature. 
The 90\% C.L. Bayesian upper limits,
computed with a uniform prior and assuming a Gaussian likelihood
function, are shown in Fig.~\ref{fig:limitComb} as a function of mass
$m_\cpoddhiggs$. The limits range from $0.7\times10^{-6}$ (at
$m_\cpoddhiggs=3.0$~GeV) to $31\times10^{-6}$ (at
$m_\cpoddhiggs=7.6$~GeV). These results are preliminary. 

\section{ACKNOWLEDGMENTS}
\label{sec:Acknowledgments}

% Standard acknowledgments paragraph; must always be included.
We are grateful for the 
extraordinary contributions of our \pep2\ colleagues in
achieving the excellent luminosity and machine conditions
that have made this work possible.
The success of this project also relies critically on the 
expertise and dedication of the computing organizations that 
support \babar.
The collaborating institutions wish to thank 
SLAC for its support and the kind hospitality extended to them. 
This work is supported by the
US Department of Energy
and National Science Foundation, the
Natural Sciences and Engineering Research Council (Canada),
the Commissariat \`a l'Energie Atomique and
Institut National de Physique Nucl\'eaire et de Physique des Particules
(France), the
Bundesministerium f\"ur Bildung und Forschung and
Deutsche Forschungsgemeinschaft
(Germany), the
Istituto Nazionale di Fisica Nucleare (Italy),
the Foundation for Fundamental Research on Matter (The Netherlands),
the Research Council of Norway, the
Ministry of Education and Science of the Russian Federation, 
Ministerio de Educaci\'on y Ciencia (Spain), and the
Science and Technology Facilities Council (United Kingdom).
Individuals have received support from 
the Marie-Curie IEF program (European Union) and
the A. P. Sloan Foundation.

% Specific acknowledgments for this paper; remove if not needed.
We wish to acknowledge Adrian Down, Zachary Judkins, and Jesse Reiss
for initiating the study of the physics opportunities with the single
photon triggers in \babar. We thank Radovan Dermisek, Jack Gunion, and 
Miguel Sanchis-Lozano for stimulating discussions.


\begin{thebibliography}{99}

\bibitem{ref:SM}
S.\ Weinberg, \jprl{19}, 1264 (1967);
A.\ Salam, p.\ 367 of {\em Elementary Particle Theory}, 
ed.\ N.\ Svartholm (Almquist and Wiksells, Stockholm, 1969);
S.L.\ Glashow, J.\ Iliopoulos, and L.\ Maiani, \jprd{2}, 1285 (1970).

\bibitem{Barate:2003sz}
LEP Working Group for Higgs boson searches, R.\ Barate \etal, 
\jpl{B565}, 61 (2003). 

\bibitem{LEPSLC:2005ema}
LEP-SLC Electroweak Working Group, Phys. Rept. {\bf 427}, 257 (2006). 

\bibitem{ref:SUSY}
J.\ Wess and B.\ Zumino, \np{B70}, 39 (1974).

\bibitem{Dermisek:2005ar}
R.\ Dermisek and J.F.\ Gunion, \jprl{95}, 041801 (2005). 

\bibitem{Dermisek:2006}
R.\ Dermisek and J.F.\ Gunion, \jprd{73}, 111701 (2006).

\bibitem{Wilczek:1977zn}
F.\ Wilczek, \jprl{39}, 1304 (1977). 

\bibitem{Dermisek:2006py}
R.\ Dermisek, J.F.\ Gunion, and B.\ McElrath, \jprd{76}, 051105 (2007). 

\bibitem{CLEO:1994ch}
CLEO Collaboration, R.\ Balest \etal, \jprd{51}, 2053 (1995). 

\bibitem{PDBook}
Particle Data Group, W.-M. Yao \etal, 
J. Phys. G \textbf{33}, 1 (2006). 

\bibitem{ref:Lozano}
E.\ Fullana and M.A.\ Sanchis-Lozano, \plb{653}, 67 (2007). 

\bibitem{ref:etab}
\babar\ Collaboration, B.\ Aubert {\em et al.},
preprint arXiv:0807.1086 [hep-ex], accepted to Phys.\ Rev.\ Lett.

% The NIM detector performance paper
\bibitem{detector}
\babar\ Collaboration, B.\ Aubert {\em et al.},
\nima{479}, {1} (2002).

\bibitem{geant}
GEANT4 Collaboration, S. Agostinelli {\it et al.},
\nima{506}, 250 (2003).

\bibitem{ref:LAT}
ARGUS Collaboration, A.\ Drescher \etal, 
\nima{237},  464 (1985).

\bibitem{ref:CBshape}
M.~J.~Oreglia, Ph.D Thesis, report SLAC-236 (1980), Appendix D;
J.~E.~Gaiser, Ph.D Thesis, report SLAC-255 (1982), Appendix F;
T.~Skwarnicki, Ph.D Thesis, report DESY F31-86-02(1986), Appendix E.

\end{thebibliography}
\end{document}